\documentclass[aps, pre, reprint, secnumarabic]{revtex4-2}

\usepackage[left=2.5cm,right=2.5cm, top=3cm,bottom=3cm]{geometry}
\usepackage[english]{babel}
\usepackage{amsmath,amssymb,amsthm,mathtools,bm}
\usepackage{euscript,mathrsfs}
\usepackage{graphicx}
\usepackage{wrapfig}
\usepackage[dvipsnames]{xcolor}
\usepackage{indentfirst}
\usepackage{amsbsy}
\usepackage{cancel}
\usepackage{verbatim}
\usepackage{empheq}
\usepackage{adjustbox}
\usepackage[pdfencoding=auto]{hyperref}
\definecolor{urlcolor}{HTML}{990000}
\definecolor{linkcolor}{HTML}{005F5F}
\hypersetup{pdfstartview=FitH,  linkcolor=linkcolor,urlcolor=urlcolor, colorlinks=true,citecolor=blue}

\renewcommand{\phi}{\varphi}
\renewcommand{\epsilon}{\varepsilon}

\DeclareMathOperator{\1}{\mathbf{1}}

\newcommand{\iu}{\mathrm{i}}

\newcommand{\E}{\mathbb{E}}
\newcommand{\R}{\mathbb{R}}
\newcommand{\T}{\mathbb{T}}

\newcommand{\Z}{\mathbb{Z}}
\newcommand{\N}{\mathbb{N}}
\newcommand{\Prob}{\mathbb{P}}

\newcommand{\bW}{\boldsymbol{W}}
\newcommand{\bK}{\boldsymbol{K}}

\newcommand{\cF}{\mathcal{F}}
\newcommand{\cL}{\mathcal{L}}

\usepackage[dvipsnames]{xcolor}
\usepackage{tikz-cd}
\usepackage{tikz}
\usetikzlibrary{positioning}
\usetikzlibrary{calc}

\definecolor{mylightred}{RGB}{211,79,73}
\definecolor{mydarkred}{RGB}{199,44,38}
\definecolor{mylightgreen}{RGB}{78,153,67}
\definecolor{mydarkgreen}{RGB}{43,129,33}
\definecolor{mylightpurple}{RGB}{150,107,178}
\definecolor{mydarkpurple}{RGB}{126,78,160}
\definecolor{mylightblue}{RGB}{49,101,205}
\definecolor{mydarkblue}{RGB}{20,92,205}

\tikzset{
  juliadot/.style args={#1,#2}{shape=circle,line width=0.03ex,minimum width=0.4ex,fill=#1,draw=#2}
}

\begin{document}


\title{Graphon Spin Systems as Exactly Solvable Models}
\author{Artem Alexandrov}
\email{alexandrov.aa@mi-ras.ru}
\affiliation{Steklov Institute of Mathematics, Moscow, Russia}
\affiliation{Phystech School of Applied Mathematics and Computer Science, Moscow Institute of Physics and Technology, Dolgoprudny 141700, Russia}
\affiliation{HSE University, Moscow, Russia}

\author{Georgi Medvedev}
\email{medvedev@drexel.edu}
\affiliation{Department of Mathematics, Drexel University, 3141 Chestnut Street, Philadelphia, PA 19104}

\begin{abstract}
  Graphons are measurable functions used to describe the asymptotic behavior of convergent graph
  families. Originally motivated by problems in combinatorics and graph theory, graphons have found
  numerous applications in the modeling and analysis of dynamical processes on networks.

  In this work, we use graphons to formulate the Ising model on convergent graph sequences, which
  include many network topologies common in applications. We derive
  the mean-field limit
  for the resulting model and obtain exact results for phase transitions in such systems.
  Specifically, we show that the critical temperatures of the Ising model on graphons are
  determined by the eigenvalues of the Hilbert--Schmidt operator associated with the graph limit.
  For many important network topologies, these eigenvalues can be computed explicitly.

  We illustrate our results with three representative random network models:
  Erd\H{o}s--R{\' e}nyi, small-world, and power-law. In the small-world case, we
  demonstrate phase transitions to both ferromagnetic and antiferromagnetic phases, as well as
  coexistence of local minima of the free energy. The latter gives rise to multistability, as confirmed
  by Monte Carlo simulations.

  The results of this work demonstrate that the Ising model on graphons combines the
  analytical tractability of exactly solvable mean-field models with the ability to accommodate a
  broad range of network topologies.
  We expect that the use of graphons in spin models will lead to new insights into the statistical physics
  of interacting systems on complex networks.
\end{abstract}

\maketitle

\section{Introduction}

The Ising model has served as a paradigmatic model for phase transitions, critical phenomena, and
collective behavior in interacting particle systems \cite{FriVel-StatMech, Ligg-IntPart}. Its influence extends
far beyond ferromagnetism, its original context, encompassing the theory of spin glasses and disordered
systems \cite{sherrington1979spin, panchenko2012sherrington, parisi2023nobel},
polymer physics \cite{nahum2013loop}, neural networks \cite{pei2025, barney2024, kropff2005storage},
and models of opinion dynamics
and social networks \cite{lynn2016maximizing, galam1997rational, brock2001discrete},
to name a few areas of applications.

Since its introduction in 1925 \cite{Isi1925}, the Ising model has been studied extensively and remains an
active area of research \cite{mattis1976, coja2019spin, coja2022ising, cugliandolo2020mean,
ostilli2024exact, dembo2010ising, dumer2023ising, bartolozzi2006spin, mezard2006reconstruction,
DemMont2010, macy2024ising}. Beyond its numerous applications, the Ising model has also
stimulated the development of new mathematical and computational techniques,
which had a profound impact on probability and statistics
\cite{Talagrand2011a, Talagrand2011b, Chat2007},
computer science, combinatorics, and optimization \cite{MezMont-book, Mont2026}.

In the Ising model an atom located at site $i\in V$ is assigned a spin $\sigma_i\in \{1,-1\}$.
Here, $V$ stands for the set of sites.
For a given spin configuration $\sigma=(\sigma_i,\, i\in V)$, the energy of the system is
given by the Hamiltonian
\begin{equation}\label{Ham}
H(\sigma)=-\sum_{i,j \in V} J_{ij} \sigma_i\sigma_j - h \sum_{i\in V}\sigma_i,
\end{equation}
where $J_{ij}$ are interaction weights and $h$ is the external field.
For simplicity, we set $h\equiv 0$ for the remainder of this paper.

In the original Ising model, spins are placed on a lattice with
nearest-neighbor interactions, i.e.,
$$
J_{ij}\neq 0
$$
only when vertices $i$ and $j$ are adjacent (denoted $i\sim j$).
This setting models magnetic interactions on a crystal lattice.

Several alternative interaction types have also been extensively studied.
In the Curie--Weiss model,
$$
J_{ij}=n^{-1}, \qquad i\neq j,
$$
so that every spin interacts equally with every other spin. Replacing nearest-neighbor couplings by
all-to-all mean-field interactions simplifies the analysis and provides an analytically tractable
framework for understanding phase transitions in the Ising model
\cite{FriVel-StatMech, BovKur2009}.

Another important example of a mean-field model is the Sherrington--Kirkpatrick model, in which
$$
J_{ij}=g_{ij}n^{-1/2},
$$
where $(g_{ij}, \,i,j\in V)$ are independent
(modulo the symmetry constraint $g_{ij}=g_{ji}$)
Gaussian random variables \cite{sherrington1979spin}.
This model plays a central role in the theory of spin glasses \cite{BinYou1986}.

The Ising model provides a unifying framework for studying collective phenomena across the physical,
biological, and social sciences. Given its broad range of applications, there has been considerable
interest in understanding the behavior of the model on general graphs, a problem that naturally
falls within the scope of network science \cite{DorMen-book}. The Ising model has been studied on
tree-like graphs \cite{dembo2010ising}, Erd\H{o}s--R{\' e}nyi random graphs
\cite{EllisNewman1978, BovGay1993, DemMont2010}, small-world graphs \cite{dumer2023ising},
and graphs with power-law degree distributions \cite{Leone2002, DomGia2010}.

A central question for the Ising model on networks is how the topology of the underlying graph affects
the thermodynamic properties of the system. The answer generally depends on structural characteristics of
the graph family under consideration, e.g.,  its degree distribution or edge density.
As a result, the analysis of the Ising models on different graphs  is often specific to a particular family
of graphs, and frequently
involves heuristic considerations or relies on numerical simulations.

A similar situation existed in the study of synchronization phenomena, where until recently rigorous results
were
available only for a few special, typically highly symmetric, network topologies. This changed with the
introduction
of graphons (a.k.a. graph limits), measurable functions that describe the asymptotic behavior of convergent
graph sequences \cite{Lovasz-book}. Graphon-based methods have since led to substantial advances
in the analysis
of the Kuramoto model of coupled phase oscillators
\cite{Med2014a, KVMed2018, CMM2018, ChiMed19a, MedMiz22,ChiMed22,CMM23},
a paradigmatic system for studying synchronization in complex networks.

The goal of this paper is to demonstrate that graphons provide an equally fruitful
framework for the analysis
of spin systems. In particular, we show that the use of graph limits leads to a unified treatment of the
Ising model on
a broad class of graphs and allows one to relate its thermodynamic behavior directly to the spectral properties
of the graph limit.

Graphons have recently been used by Searle and Tindall to study spin systems on random graphs
\cite{SeaTin2024}.
In that work, both a quantum system of interacting qubits and a classical Ising model were considered.
For the latter,
the authors derived the thermodynamic limit and established the existence of phase transitions.
Their analysis also
yielded explicit solutions of the mean-field equation for certain rank-one graphons. In contrast, the approach
developed in the present paper applies to general graphons and relates the critical temperatures directly to
the spectral properties of the self-adjoint Hilbert--Schmidt operator associated with the graph limit.
For many graph families of interest, the eigenvalues of this operator can be computed explicitly.

Specifically, in the present paper, we derive explicit analytical expressions for the critical temperatures
and relate them
to spectral characteristics of the graph limit and, consequently, to the asymptotic connectivity of the
underlying network.
We analyze the Ising model on three representative network topologies:
Erd\H{o}s--R{\' e}nyi, power-law,
and small-world. Our results for Erd\H{o}s--R{\' e}nyi and power-law networks are consistent with
those of previous studies \cite{EllisNewman1978, BovGay1993, SeaTin2024}. Our findings
for small-world networks appear to be new. This case is of particular interest because small-world
organization has been observed in a wide variety of real-world networks
(see, e.g., \cite{WatStr1998, BassettBullmore2006}).

Since our primary interest lies in understanding the relationship between network structure
and system behavior, we restrict to simple graphs. However, the continuum limit developed
below extends
naturally to more general classes of weighted directed sparse random graphs
(cf., \cite{Med2014a, Med2014b, Med19}).

The organization of the paper is as follows. In the next section, we formulate the Ising model
on graphs. In Section~\ref{sec.Wrandom}, we explain the construction of $W$-random graphs
and discuss in detail three representative families of $W$-random graphs: Erd\H{o}s-R{' e}nyi,
small-world, and power-law. We then derive the mean-field limit for the Ising model on $W$-random
graphs and analyze the phase transitions of the continuum limit in Section~\ref{sec.mean-field}.
In Section~\ref{sec.examples}, we discuss phase transitions on the three networks:
Erd\H{o}s-R{' e}nyi, small-world, and power-law. For small-world networks, we show that
there are countably many phase
transitions of the paramagnetic phase leading to coexisting local minima of the free energy.
The coexisting solutions of the mean-field equation give rise to multistability, which can be
observed in Monte-Carlo simulations summarized in Section~\ref{sec.Monte-Carlo}. We conclude this
paper with a brief discussion in Section~\ref{sec.discuss}.

\section{The model}\label{sec.model}

In this section, we formulate the Ising model on $W$-random graphs and
derive the thermodynamic limit.

Consider the Ising model with the Hamiltonian \eqref{Ham}. As the
interaction weights, we take $J_{ij}=n^{-1}J a_{ij}$, $i,j\in [n]\doteq\{1,2,\dots, n\}$.
Here, $A=(a_{ij})$ is the adjacency matrix, which defines the connectivity of the network.
Since we are interested in the effects of the network connectivity on the phase transitions
in the Ising model, we restrict to simple networks, i.e., $A$ is a symmetric $\{0,1\}$-valued
matrix with zeros on the main diagonal. As will be clear below $A$ does not have to be symmetric.
The results below naturally apply 
to a more general class of convergent graph sequences, including sparse directed weighted
graphs (cf.~\cite{Med19}).

Further, $J$ can be either positive or negative depending on the type of the interactions.
$J>0$ corresponds to ferromagnetic regime, while $J<0$ corresponds to antiferromagnetic.
Interactions of both types are also used in modeling systems outside the field of magnetism.
For instance, in the context of opinion dynamics or consensus protocols $J$ of different
signs may correspond to cooperative versus noncooperative dynamics.

There is no difficulty, at least at the modeling level, in allowing $J_{ij}$ to take
values of both signs, as in the Sherrington--Kirkpatrick model \cite{BovKur2009}. However,
the analysis of such models requires additional care. To keep the presentation simple and to
focus on the main features of the graphon-based approach, in this paper, we restrict to models
with a fixed interaction type, namely, either ferromagnetic or antiferromagnetic interactions.

The first step in the analysis of the Ising model is the derivation of the free energy. The main
reason why the analysis of the mean-field Curie-Weiss model is much simpler than the analysis
of the classical Ising model with nearest neighbor coupling is that the free energy
can be expressed in terms of the magnetization $m=n^{-1}\sum_{i\in [n]} \sigma_i$.

While this approach cannot be applied directly to a general structured network, for dense
networks it can be suitably modified leading to a significant simplifications in the analysis.
The key step is to replace the underlying network by a blow-up network (cf.~\cite{BCLV2008}).
Specifically, we replace each node of the graph by an all-to-all connected
cluster of $\nu$ nodes. This will allow us to operate with locally averaged variables
such as local magnetization. This effectively allows to apply the mean-field approach to structured
networks.

Formally, at  each node~$i\in[n]$, we place $\nu\gg 1$ all--to--all coupled spins
$\sigma_i^s\in\{-1, 1\}, s\in [\nu]$. The Hamiltonian then becomes
\begin{equation}\label{pre-Ham}
  H(\sigma)=-\frac{1}{2 n\nu^2}\sum_{i,j=1}^n\sum_{s,t=1}^\nu J_{ij} \sigma_i^s\sigma_j^t.
\end{equation}

As advertised above, the local magnetization is given by 
$$
m_i\approx \nu^{-1}\sum_{s=1}^\nu\sigma_i^s,
$$
where the approximate equality becomes the exact equality
when $\nu\to\infty$.

The free energy is then expressed as
$$
F(m)=\frac{-1}{2n}\sum_{i,j=1}^n J_{ij} m_im_j -T\sum_{i=1}^nS(m_i),
$$
in exact analogy to the free energy in the exactly solvable Curie-Weiss model \cite{BovKur2009}.
Here, the entropy
$$
S(x)=-\ell\left(\frac{1+x}{2}\right)-\ell\left(\frac{1-x}{2}\right),\; \ell(x)=x\log x.
$$

After setting $\partial_{m_i}F(m)=0$, we arrive at
\begin{equation}\label{steady-i}
  m_i=\tanh\left( \beta n^{-1}\sum_{j=1}^n a_{ij} m_j\right), \; i\in [n],
\end{equation}  
where we used $J_{ij}=Ja_{ij}$ and $\beta=JT^{-1}$. In case of the directed network (i.e., if $A$
is not symmetric), $A=(a_{ij})$ is replaced by its symmetric part, which means that in
\eqref{steady-i}, $a_{ij}$ is replaced $\frac{1}{2}\left(a_{ij}+a_{ji}\right)$. Thus,
as a bonus, our approach automatically covers models with directed interactions.

\section{W-random graphs}\label{sec.Wrandom}
To ensure that the mean-field limit in Eq.~\eqref{steady-i} exists, the graph sequence
\(\Gamma_n\) must converge in the sense of convergence of dense graphs (cf.~\cite{Lovasz-book}).
In principle, for the results presented below to hold we only need to assume convergence
of \( \Gamma_n\) in cut-norm topology. To simplify the presentation, we take a special class
of convergent graph sequence given by $W$-random graphs. It is a representative random graph
model, which includes many interesting graphs important for applications.

To construct a $W$-random graph,  we need a symmetric
measurable function on a unit square. Let $W: Q \times Q \to [0,1]$, where $Q$
stands for the unit interval $[0,1]$.
$W$ is called \textit{a graphon} in the theory of graph limits,
\cite{Lovasz-book}. It
describes a limiting behavior of the graph sequence, which we are
about to generate. Graphons in the form of step-functions are also used  
to represent  finite graphs.

Next, we partition $Q$ into $n$ subintervals $Q_i^n = [(i-1)/n, i/n)$ and average $W$ over
$Q_{ij}^n = Q_i^n \times Q_j^n$:
\begin{equation}\label{def-Wn}
W^n_{ij}=n^2\int_{Q^n_{ij}} W(x,y)dxdy, \; i,j\in [n].
\end{equation}
At this step, we have defined a step-function
$$
W^n=\sum_{i,j=1}^n W^n_{ij} \1_{Q_i^n}(x) \1_{Q_j^n}(y).
$$

The entries of the adjacency matrix $a_{ij}, 1\le i\le j\le n,$ are defined as independent binary
random variables with
\begin{equation}\label{def-Wrandom}
  \Prob(a_{ij}=1)=W^n_{ij}, \quad \Prob(a_{ij}=0)=1-W^n_{ij}.
\end{equation}

This yields a random $\{0,1\}$-valued graphon
\begin{equation}\label{def-X}
X^n=\sum_{i,j=1}^n W^n_{ij} \1_{Q_i^n}(x) \1_{Q_j^n}(y).
\end{equation}

It is instructive to review the roles the three graphons $X^n$, $W^n$,
and $W$ play in the modeling of interacting spin/particle models and
the analytic relationships between them.

The random graphon $X^n$ specifies the interactions in the interacting spin
model on a random network. $W^n=\E(X^n)$, the expected value of $X^n$,
is used to define a deterministic averaged model on a finite network,
while $W$ represents the connectivity in the continuum limit as the size of networks
tends to infinity.

The key for justifying the continuum limit is the fact that $\|X^n- W^n\|_\Box\to 0$
as $n\to\infty$ with probability $1$. This is a manifestation of the Law of Large Numbers
for graphons. The Large Deviation Principle for graphons was proved in \cite{DupMed2022}.
The cut-norm is defined as follows
\begin{equation}\label{cut}  
  \|U\|_\Box=\sup_{A\times B} \left| \int_{A\times B} U(x,y) dx\,dy\right|,
\end{equation}
where $A$ and $B$ are measurable subsets of $[0,1]$. The cut-norm metrizes
graph convergence \cite{Lovasz-book}.

On the other hand, $W^n$ tends to $W$ almost surely and in $L^1$-norm by the standard
results in analysis (see \cite{Chat2017} for an accessible discussion of the graph limit theory
and related facts from analysis).
These analytical considerations were used to justify the continuum limit for coupled dynamical models
on random graphs \cite{Med2014b, Med19, DupMed2022}. The same strategy can be used to
justify the continuum limit for the Ising model on random graphs that will be introduced below.

\begin{figure*}[t]
    \begin{center}
    {\bf a}\,\includegraphics[width=0.31\linewidth]{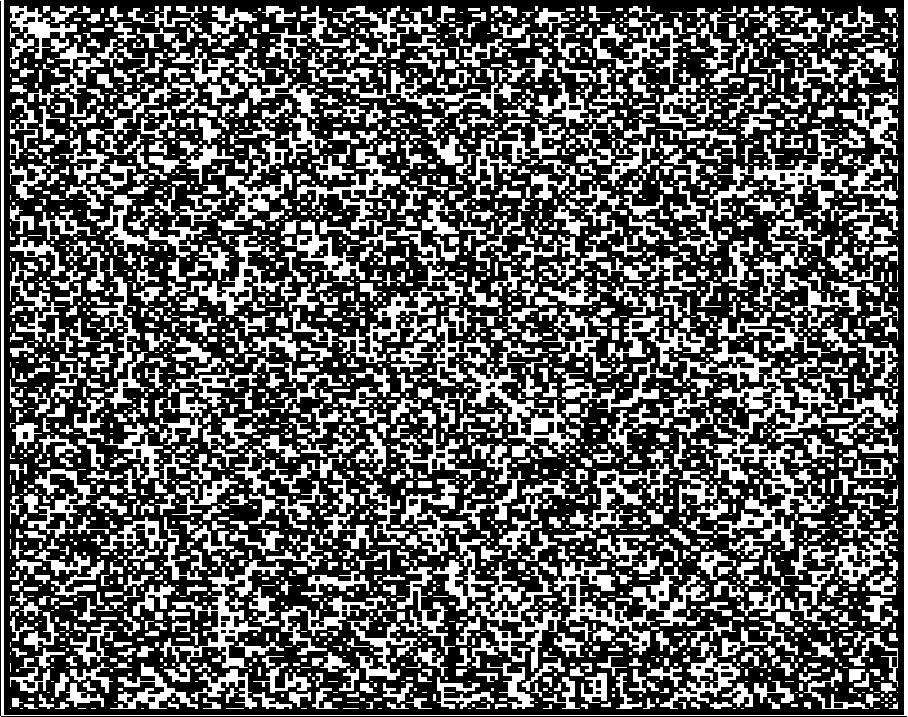}
   \hfill
   {\bf b}\,\includegraphics[width=0.31\linewidth]{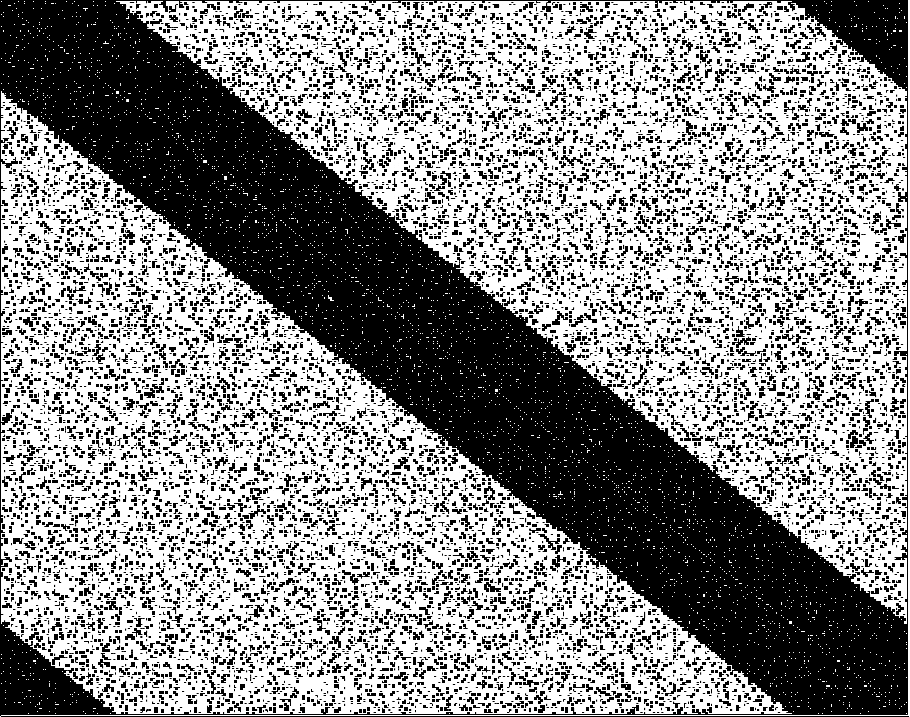}
   \hfill
   {\bf c} \includegraphics[width=0.31\linewidth]{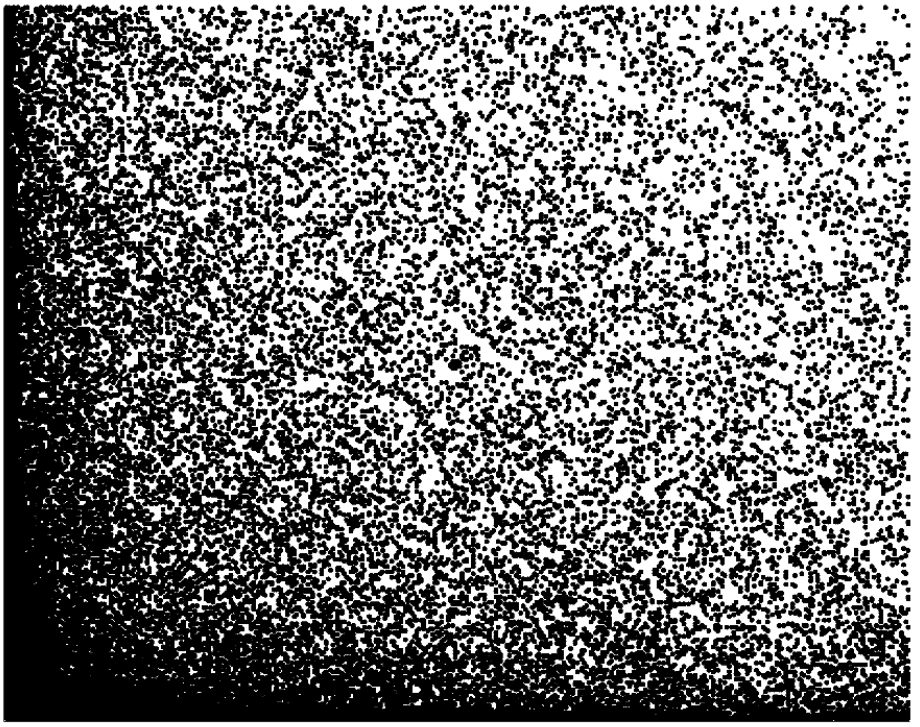}
    \end{center}
    \caption{Pixel pictures of Erd\H{o}s-R{\' e}nyi, small-world, and power-law
      graphs.}
    \label{f.pixelER}
  \end{figure*}

We will now discuss three representative examples of $W$-random graphs that will be used below:
 Erd\H{o}s-R{\' e}nyi, small-world, and power-law graphs (see Fig.~\ref{f.pixelER}).
We begin with the classical Erd\H{o}s-R{\' e}nyi graph model \cite{Bollob-RandomGraphs}. 
To this end let $W\equiv p\in (0,1)$.  Fig.~\ref{f.pixelER}(\textbf{a}) shows a pixel picure of the
$W$-random graphs with constant graphon $W\equiv 1/2$ and 
$n=600$. The almost uniform distribution of pixels in this figures provides a
geometric intuition for the claim that the limiting behavior of a given random graph sequence
is captured by the constant graphon $W$, which
is equal to the density of pixels per unit area.
Plots in \textbf{b} and \textbf{c}
show similar pixel pictures of $W$-random graphs corresponding to small-world and power-law
graphons that will be explained below.

The Ising model on the  Erd\H{o}s-R{\' e}nyi
graph is a random
counterpart of the Curie-Weiss model. One can show that with probability $1$ all results for the
Curie-Weiss model hold for the Ising model on  Erd\H{o}s-R{\' e}nyi graphs.
$W$-random graphs generalize this idea much further. With the appropriate
choice of $W$, we can model a great variety of graphs, which locally look like an Erdos-Renyi
graph, and thus can be treated by suitably adjusting the mean-field approach, which was so
effective for the Curie-Weiss model (cf.~\cite{BovKur2009}).

\begin{figure*}[t]
    \centering
    {\bf a}\,\includegraphics[width=0.31\textwidth]{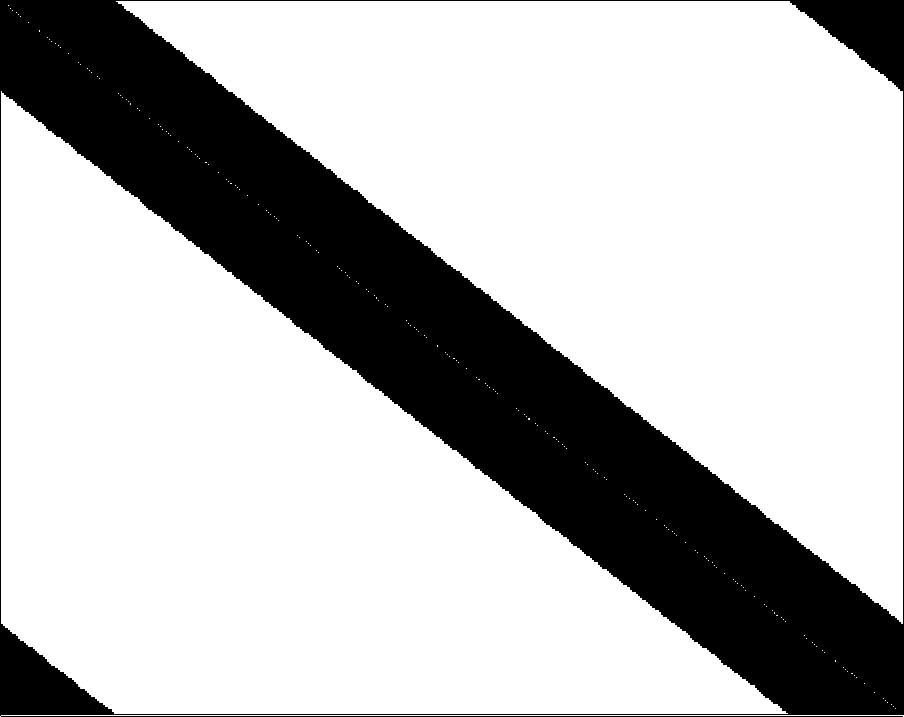}
    \hfill
    {\bf b}\,\includegraphics[width=0.31\textwidth]{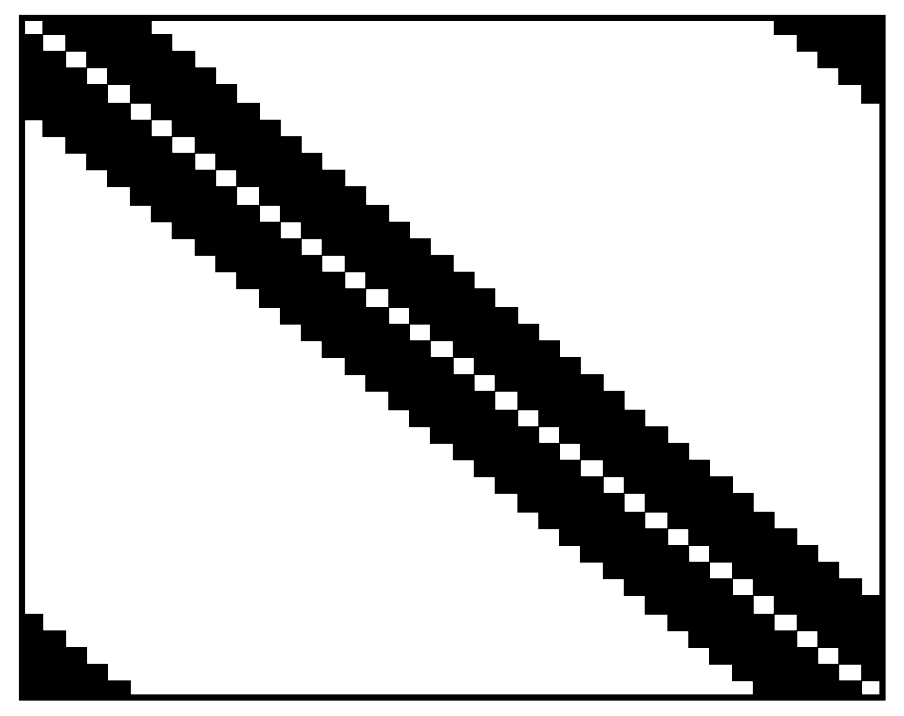}
    \hfill
    {\bf c}\,\includegraphics[width=0.31\textwidth]{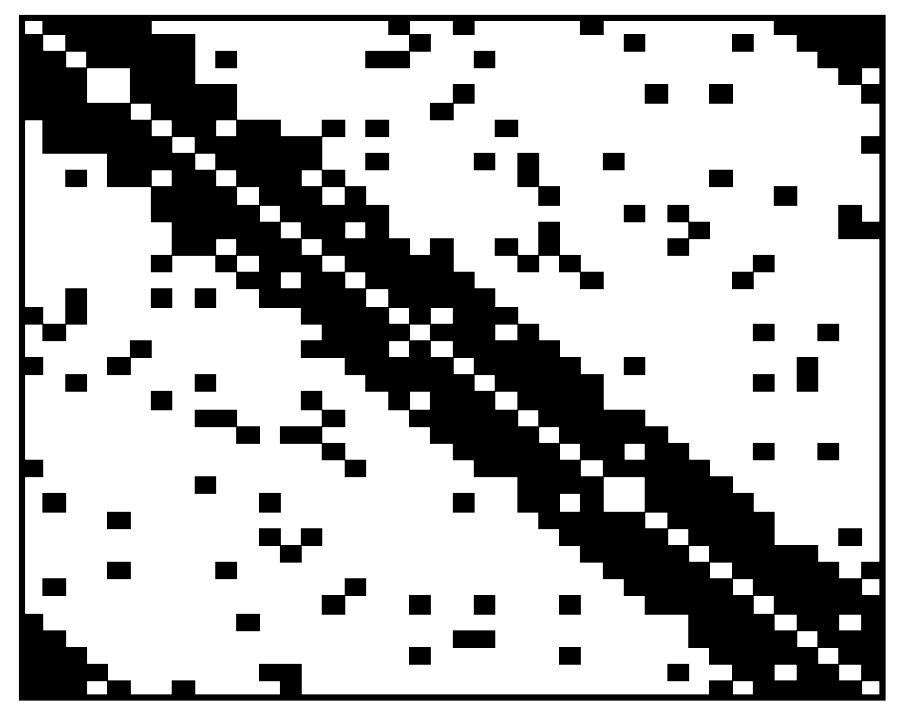}
  \caption{Construction of a small-world network. Starting with a graphon $W$ (\textbf{a}),
      construct the deterministic step function $W^n$ determining the probabilities of
      connections (\textbf{b}), and generate a random matrix $A$ (\textbf{c}).}
    \label{f.small-w}
  \end{figure*}

To demonstrate the power of graphons in generalizing the mean-field
Curie–Weiss model to structured networks,
we consider the Ising model on a family of small-world graphs.
To construct a small-world graph, let 
$ W(x,y) = K(x-y)$:
\begin{equation}\label{sw-kernel}
 K(x)=
  \begin{cases}
    1-p, & |x| \le r, \\
    p, & r < |x| \le 1/2,
  \end{cases}
\end{equation}
where $K$ is a symmetric function defined on $[-1/2,1/2]$ and extended to
$\R$ by periodicity. By averaging this function over discrete cells $Q_{ij}$, we
first obtain step function $W^n$ (cf.~\eqref{def-Wn}), which is then used to define the random
adjacency matrix $A^n$ (cf.~\eqref{def-Wrandom}). The construction of the small-world graph
is illustrated in Fig.~\ref{f.small-w}. The pixel picture in plot {\bf c} shows the distribution
of edges in a random small-world network. The sequence of $W$-random small-world graphs
$\Gamma^n$ converges in cut-norm to the original graphon $W$ with probability $1$
\cite{DupMed2022}. Convergence in cut-norm in particular implies that the eigenvalues and the
corresponding eigenspaces of the kernel operators
 \begin{equation}\label{kernel-Wn}
 \bW^n[u]\doteq \int_Q W^n(\cdot,y)u(y)~dy
 \end{equation}
 converge to the eigenvalues and the eigenspaces of the limiting kernel operator
\begin{equation}\label{kernel-W}
 \bW[u]\doteq \int_Q W(\cdot,y)u(y)~dy
\end{equation}
(cf.~\cite{GhaMed2025, Sze2011}).
This fact will be crucial for the analysis of  phase transitions in the Ising model on large
small-world networks. It, in particular, implies that the analysis of the phase transitions
in large (finite) networks can be reduced to the analysis of the continuum limit
based on the limiting structure of the network encoded in $W$. Note that although the
kernel operators representing
individual graphs $\bW^n$ are random, the limiting operator $\bW$ is a deterministic
\textit{shift-invariant} Hilbert--Schmidt operator whose eigenvalues can be computed explicitly
using the Fourier transform \cite{Med2014c}. This yields a major advantage of the continuum limit
in the random setting, as it results in a simpler network possessing additional structural
features (like shift invariance for the model at hand) that are not present in the original
discrete random network.

 Our final example shows that the continuum limit procedure applies to sparse
networks as well. Many real-world networks are sparse. At the same time, sparse networks
are harder to analyze. Fortunately, the $L^p$-graphon
framework extends the graph limit theory to cover sparse networks \cite{borgs2019Lp1}.
The following example of sparse power-law graphs illustrates this case
(cf.~\cite{MedTan2018}).

  Let $0<\alpha<1/2$ and
  \begin{equation}\label{W-power}
    W(x,y)= (xy)^{-\alpha}.
    \end{equation}
    Unlike  graphons, which we used to define
    Erd\H{o}s-R{\' e}nyi and small-world graphs, $W$ in \eqref{W-power} is not
    bounded by $1$.
    It has a square-integrable singularity at the origin. This is important, 
    because this singularity will allow us to generate a highly connected cluster of nodes, a so-called
    hub (see Fig.~\ref{f.pixelER}\,{\bf c}).
    Many real-world networks including the Internet and social networks feature such hubs.

    Because $W$ is not bounded by $1$, the procedure for generating $W$-random graphs
    (cf.~\eqref{def-Wn}, \eqref{def-Wrandom}) needs to be implemented with care.
    For the model at hand,
    we define
    \begin{equation}\label{PL-Wn}
    W^n_{ij}=n^{-\gamma} \int_{Q_{ij}}\min\{ n^\gamma, W(x,y)\}\;dx\,dy,
  \end{equation}
  where $\gamma\in (\alpha, 1/2)$ controls the sparsity of the network.
    Note that $W^n_{ij}$ are bounded by $1$, so they can be used to define the probabilities of
    connections
    in \eqref{def-Wrandom}.

    On the other hand, as $n\to\infty$, $W^n$ more and more closely
    reflect the structure of $W$, including
     the singularity at the origin. In particular, the probabilities of connections
    for nodes $i\ll n$ are much higher than for those that lie farther away, creating a hub
    of highly connected nodes (see Fig.~\ref{f.pixelER}\,\textbf{c}).

    Another interesting feature of the power law graphs is their sparseness. A simple computation
    shows
    that the expected degree of node $i$ is of order
    $n^{1+\alpha-\gamma} i^{-\alpha}$,
    which justifies the power-law name of this family of graphs.
    Further, the expected edge density
    is $O(n^{-\gamma})$
    (cf.~\cite{MedTan2018}). This shows that this family of graphs is much sparser
    than  dense Erd\H{o}s-R{\' e}nyi and small-world graphs, for which the edge density is $O(1)$.

    The Erd\H{o}s--R{\' e}nyi, small-world, and power-law networks discussed above illustrate the
    flexibility and analytical power of the $W$-random graph model. In what follows, we show
    that this model fits naturally into the framework of the Ising model on graphs and provides
    a convenient setting for taking the mean-field limit and analyzing phase transitions.

    \section{Mean-field limit}\label{sec.mean-field}
    We are now prepared to discuss the continuum limit for the Ising model on random graphs.
    Let $W$ be a given graphon, which generates a sequence of $W$-random graphs
    via \eqref{def-Wrandom}.

    Using random graphon $X_n$ (cf.~\eqref{def-X}), we can rewrite the equation for steady states
    \eqref{steady-i} as an integral equation
\begin{equation}\label{re-steady}
m^n(x)=\tanh\left(\beta\int_Q X^n(x,y)m^n(y)dy\right), \; x\in Q,
\end{equation}
where $m^n=\sum_{i=1} m_i\1_{Q^n_i}(x).$

By sending $n\to\infty$, we formally obtain the mean-field limit of
\eqref{steady-i}:
\begin{equation}\label{clim}
\cF(u,\beta)\doteq\tanh\left(\beta\bW[u] \right) -u(x)=0,
\end{equation}
where $\bW[u]\doteq \int_Q W(\cdot,y)u(y)~dy$.

Using convergence of $X^n$ to $W$ as $n\to\infty$, one can justify that for
 values of $\beta$, for which \eqref{clim} has a unique solution, the solutions
 of the discrete system \eqref{re-steady} converge to that of \eqref{clim} in the
 $L^2$-sense with probability $1$.
This can be done using arguments similar to those used in \cite{Med2014b, Med19, DupMed2022}
for dynamical models on networks. As we are interested in the phase transitions in
the Ising model, we actually need to know that the phase transitions
in the limiting system \eqref{clim}
approximate the phase transitions in the finite system \eqref{re-steady} for large
$n$. This is a more delicate problem. A closely related problem was studied in \cite{MedPel2024}
in the context of the Turing instability. Below we adapt the results from this work to the
analysis of the phase transitions in the Ising model on random graphs.

The trivial solution $u\equiv 0$ of \eqref{clim} (the paramagnetic phase) exists for all $\beta$.
The phase transitions in the Ising model \eqref{pre-Ham} correspond to the bifurcations
of the trivial solution of \eqref{clim}. The latter are determined by the spectrum of
the kernel operator
$\bW$ (cf.~\eqref{kernel-W})
and the symmetries in \eqref{clim}.

As a self-adjoint Hilbert-Schmidt operator on $L^2(Q)$,
$\bW$ has at most a countable number of eigenvalues with a single
accumulation point at $0$:
\begin{equation}\label{spec-W}
 \lambda_1\ge \lambda_2\ge \dots\ge \lambda_0=0\ge \dots\ge
\lambda_{-2}\ge \lambda_{-1}.
\end{equation}
We denote the spectrum of $\bW$, by $\sigma(\bW)$.

Let $\beta=\beta_0+\gamma$, $
\cF_{\beta_0}(u,\gamma)\doteq \cF(u,\beta_0+\gamma)$ and
rewrite \eqref{clim} as follows
\begin{align} \nonumber
\cF_{\beta_0}(u,\gamma)&=\cL_{\beta_0}[u]+\gamma u
                         -\beta_0^2\gamma \left(\bW[u]\right)^3+\dots\\
  \label{set-F}
&=0,
\end{align}
where $\cL_{\beta_0}[u]\doteq \beta_0\bW[u]-u$.

If 
$
\beta_0^{-1} \notin \sigma(\bW),
$ 
then $\cL$, the linearized operator of $\cF$ around $u=0$, has a bounded inverse,
and by the Implicit Function Theorem, 
$
u \equiv 0
$ 
is the unique solution of \eqref{clim} for small $|\gamma|$ \cite{Dieu-Analysis}.

If
$
\beta_0^{-1} = \lambda_k \in \sigma(\bW),
$ 
then the trivial solution of \eqref{set-F} undergoes a bifurcation. The odd symmetry 
$
\cF(-u,\gamma) = -\cF(u,\gamma)
$ 
implies that this is a pitchfork bifurcation.

The bifurcations of the trivial solution of \eqref{set-F}  correspond to the phase transitions
in the Ising model.
The eigenvalues of the kernel operator $\bW$ determine the 
critical values of the temperature at which phase transitions occur.

Even when $W$ is a nonnegative graphon, $\bW$ can
have eigenvalues of both signs, giving rise to coexistence of \textit{ferromagnetic} (FM) and
\textit{antiferromagnetic} (AFM) phases on the same network.

If the operator $\bW$ has distinct nonzero eigenvalues then there will be  multiple
phase transitions and possible
coexistence of solutions of Eq.~\eqref{clim}. This shows that in general the Ising model
on $W$-random graphs may exhibit \textit{metastability}. Below, we show that this is indeed the case
for the Ising model on small-world graphs.

\section{Examples}\label{sec.examples}

We now determine the finer structure of these phase transitions for three
representative examples of $W$-random networks:  Erd\H{o}s-R\'{e}nyi graph,
small-world, and power-law.

\subsection{Erd\H{o}s-R{\' e}nyi networks}
First, consider $W \equiv p \in (0,1]$, which corresponds to an Erd\H{o}s-R\'{e}nyi graph if $p<1$,
and to the complete graph if $p=1$.

This is a rank-$1$ graphon. $\bW$ has a simple eigenvalue
$
\lambda_1 = p.
$ 
The corresponding eigenfunction 
$
\xi_1(x) \equiv 1.
$
The orthogonal complement of the $\operatorname{span}\{\xi_1\}$ is the eigenspace of the zero
eigenvalue.

\begin{figure}[h]
    \centering
     \includegraphics[width=\linewidth]{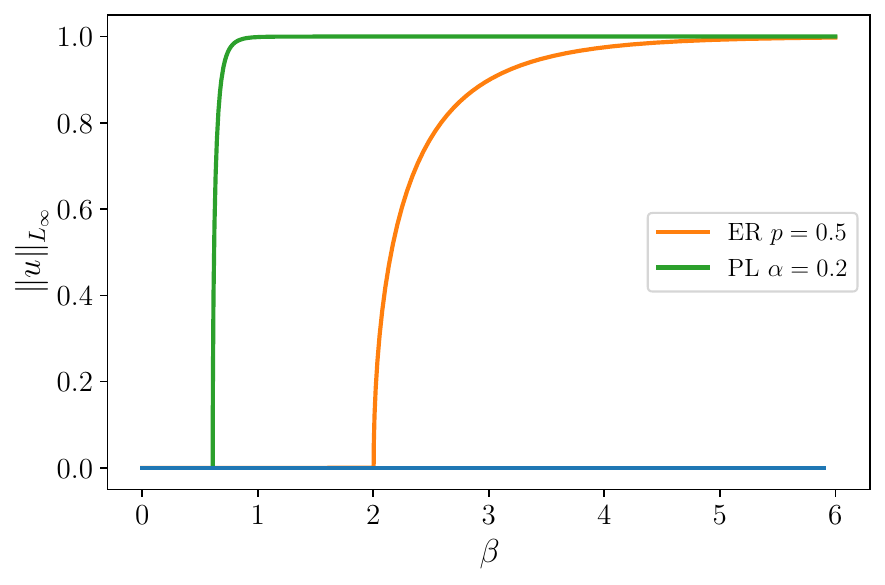}
    \caption{Bifurcations in the Ising model on Erd\H{o}s-R\'{e}nyi (ER) graphon with $p=0.5$ and power-law (PL) graphon
      with $\alpha=0.2$. In both models, $J=1$.}
    \label{f.ER}
\end{figure}

By the Crandall-Rabinowitz theorem, combined with the odd symmetry of 
\(\cF_{p^{-1}}(\cdot,\gamma)\), we conclude that at 
$
\beta = p^{-1}
$ 
a constant (nonzero) solution emerges through a pitchfork bifurcation.
We have computed the bifurcation diagram for the Ising model on Erd\H{o}s-R{\' e}nyi
graph using \verb|BifurcationKit|. Numerical bifurcation diagram
shown in Fig.~\ref{f.ER} is consistent with the analytical prediction.

\subsection{Power-law networks}
Next, we consider the power-law graphon $W=(xy)^{-\alpha}$, $\alpha \in (0,1/2)$
\cite{MedTan2018}.
Like the constant graphon $W\equiv p$, this is also a rank-1 graphon. The only nonzero eigenvalue
of $\bW$ is $\lambda_1 = (1-2\alpha)^{-1}$. The corresponding eigensubspace is
spanned by $\xi_1=x^{-\alpha}$. Thus, the paramagnetic phase undergoes a phase
transition at the inverse temperature $\beta_c = 1 - 2\alpha$. 
The numerical bifurcation diagram in Fig.~\ref{f.ER} shows a branch of nonzero solutions
emerging at $\beta_c = 1 - 2\alpha$, in agreement with our theoretical prediction.

\subsection{Small-world networks}
Next, we turn to our main example $ W(x,y) = K(x-y)$ with $K$ defined in
\eqref{sw-kernel}.


Given the periodicity of $K$, it is natural to consider \eqref{clim}
on the torus $\T \doteq \R / \Z$ instead of the unit interval $Q$.
In this setting, $\bW$ becomes a convolution operator $\bK[u]=K\ast u$
and the eigenvalues of $\bK$ are obtained by taking the
Fourier transform of $K$ (cf.~\cite{Folland-FA}):
\begin{equation}\label{eig-K}
  \mu_k=\widehat{(K)}_k\doteq\int_\T K(x)e^{-\iu 2\pi kx}dx,\; k\in\Z.
\end{equation}

\begin{figure}[h]
\begin{center}
  \includegraphics[width=\linewidth]{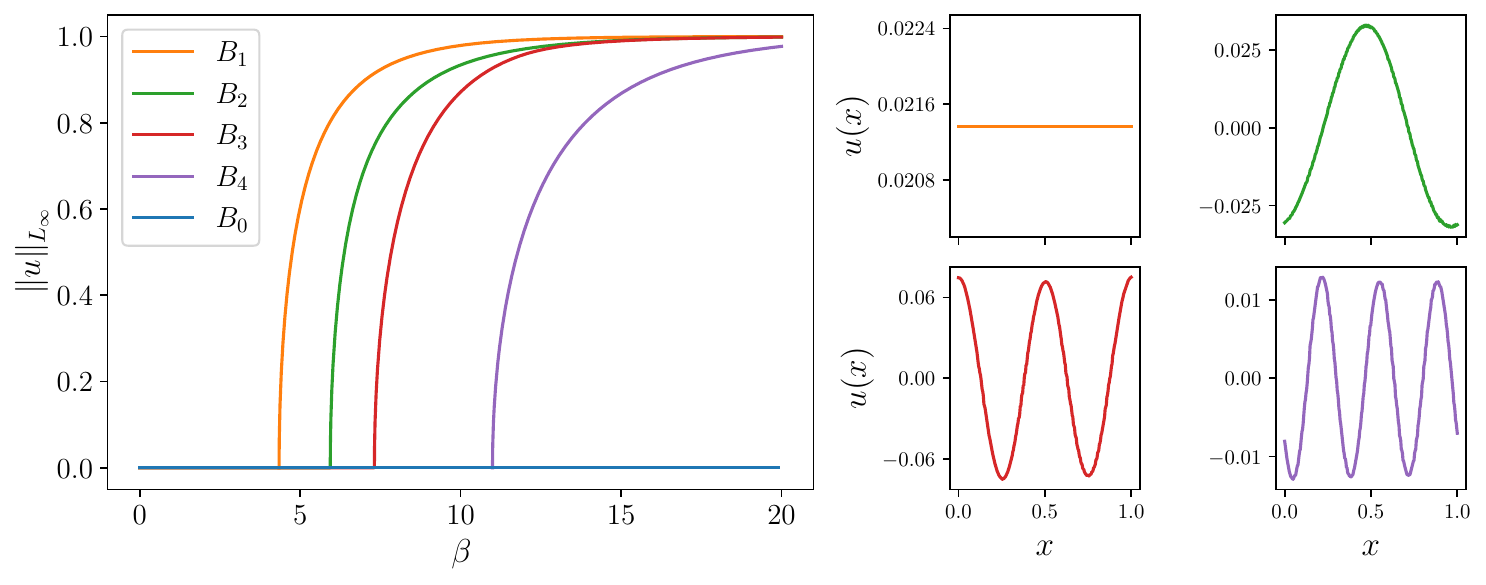}
   \includegraphics[width=\linewidth]{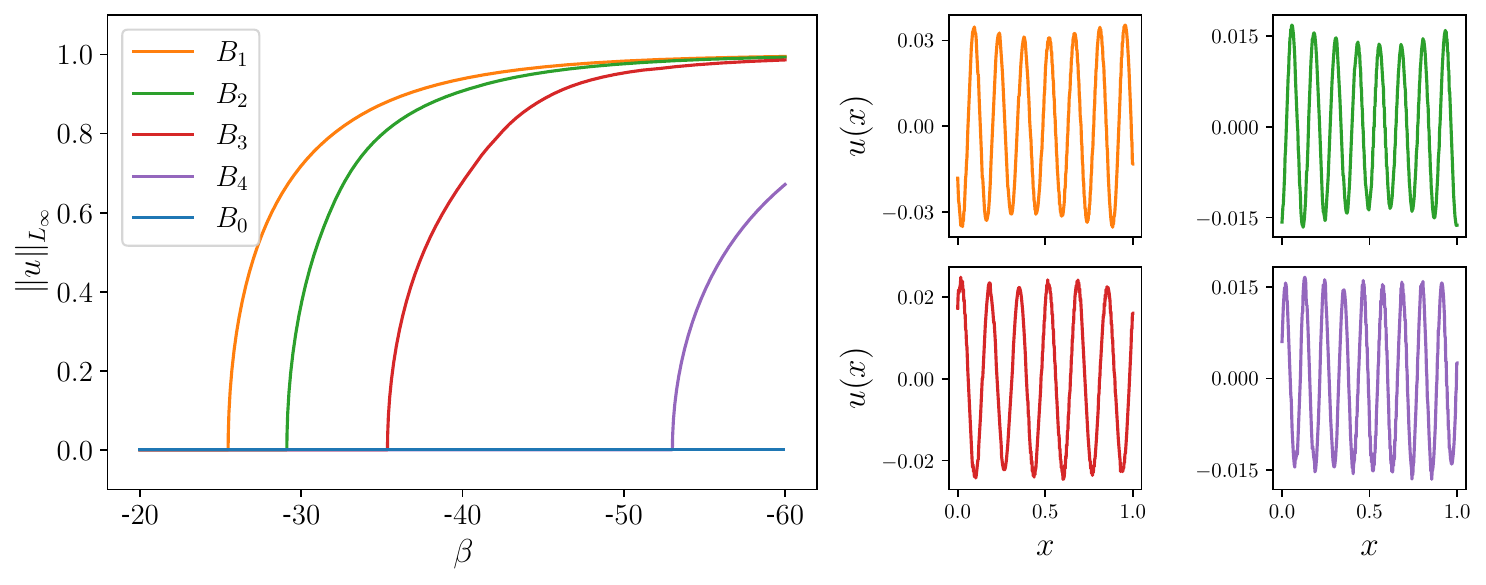}
\end{center}
\caption{Bifurcation diagrams for \eqref{clim} for the small-world
  network with $p=0.05$ and $r=0.1$. }
\label{f.bif-sw}
\end{figure}

The corresponding eigenfunctions are $\xi_k=e^{-\iu 2\pi kx}$, $k\in\Z$.
The bifurcation analysis of \eqref{clim} takes into account the following symmetries of \eqref{clim}:
  if $u(x)$ is a solution of \eqref{clim} then so are
  \begin{equation}\label{symmetries}
    u(x+h), \quad -u(x),\quad\mbox{and}\quad u(-x).
    \end{equation}
    The first follows from translation invariance of \eqref{clim}, the second from the odd symmetry
    $\cF(-u,\beta)=-\cF(u,\beta)$, and the third from the evenness
of the kernel $K(-x)=K(x)$. These symmetries determine the normal form of the
equation near the bifurcation. The pitchfork bifurcation subject to these symmetries is analyzed in detail in
\cite[\S~3]{MedPel2024}.
Here, we outline the key steps.

First, note that since $K(x)$ is even and real, $\mu_{k}=\mu_{-k}$ for all $k\in\N$.
Therefore, the multiplicity of $\mu_k$ is at least two. $\mu_0$ is simple and the corresponding
eigenfunction is constant $\xi_0\equiv 1$. Let $\beta_0=\mu^{-1}_p, p\in\N,$ and suppose the multiplicity
of $\mu_p$ is $2$. By applying the Fourier transform to \eqref{set-F}, we have
\begin{align} \nonumber 
  &\left[\frac{\mu_k}{\mu_p}\hat{u}_k - \hat{u}_k\right]  -\gamma \hat{u}_k \\
  \label{F-eqn}
  &+
\frac{\mu_k^3}{\mu_p^2} \sum_{k_1,k_2\in\Z} \hat{u}_{k_1}\hat{u}_{k_2}\hat{u}_{k-k_1-k_2}+\dots
=0,
\end{align}
for $k\in\Z$. Here, we used $\widehat{K\ast u}_k=\mu_k\hat{u}_k$, and $\beta_0=\mu_p^{-1}$.
\begin{figure}
\begin{center}
  \includegraphics[width=0.75\linewidth]{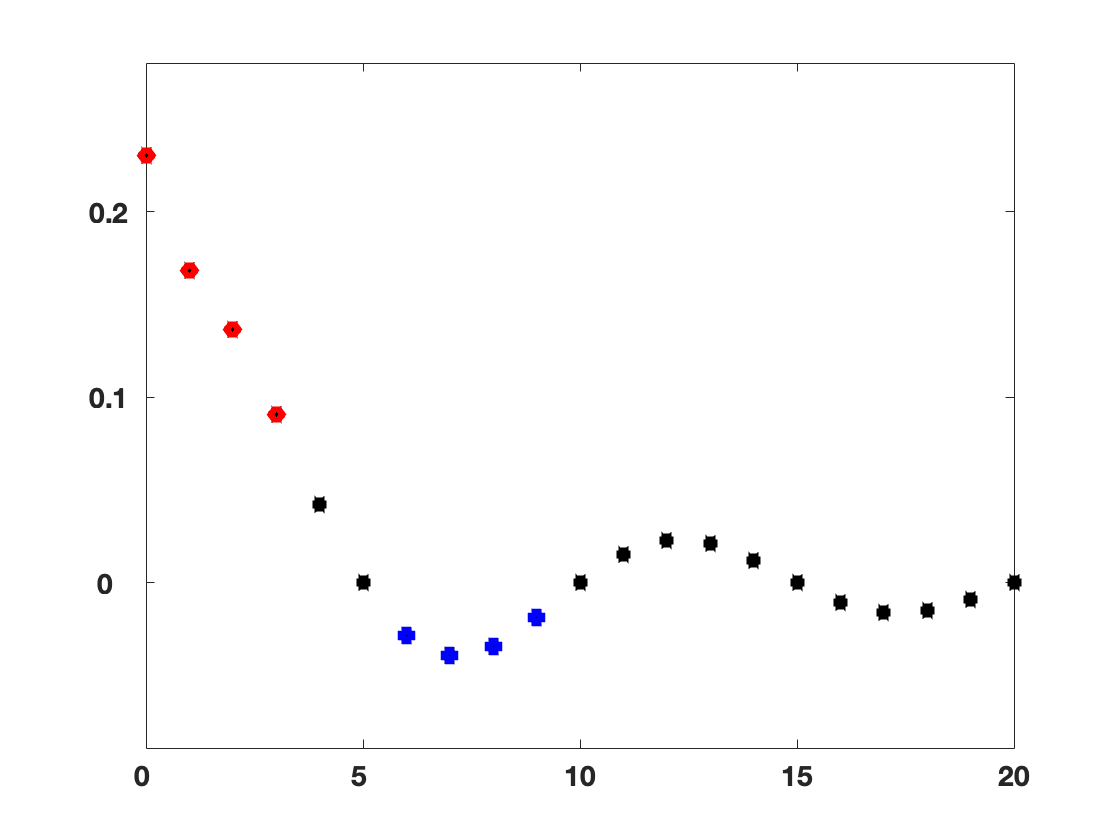}
\end{center}
\caption{Eigenvalues of $\bK$ for small-world graph with $p=0.05$ and
  $r=0.1$. The first four smallest and largest eigenvalues
are plotted in blue and red respectively.
}
\label{f.eigenval}
\end{figure}

Translational invariance of \eqref{clim} (cf.~\eqref{symmetries}) results  in SO(2) symmetry for
the Fourier coefficients in \eqref{F-eqn}. Consequently, the sum in \eqref{F-eqn} simplifies
to $3 \hat u_k |\hat u_k|^2$ ($k_1=\pm k$). The bifurcation equation is then obtained 
from the equation for $p=k$ in \eqref{F-eqn}
$$
\gamma A-3\mu_pA|A|^2+\dots=0, \; A:=\hat u_p.
$$
From here we find $|A|\approx\sqrt{ \frac{\gamma}{3\mu_p}}$ and the solution bifurcating
from $0$ at $\beta=\mu_p^{-1}$ to leading order and up to a shift is given by
\begin{equation}\label{harmonic}
u_\gamma(x)\approx \sqrt{ \frac{\gamma}{3\mu_p}}\cos\left(2\pi px\right).
\end{equation}
The case of the simple eigenvalue $\mu_0$ is analyzed similarly. The main distinction is
that the bifurcating solution is constant in contrast to harmonics in
\eqref{harmonic} for
$k\neq 0$ (see Fig.~\ref{f.bif-sw}).

The eigenvalues of $\bK$ can be computed explicitely (cf.~\cite{MedPel2024}):
\begin{equation*}
\mu_k=\left\{ \begin{array}{ll} 2r(1-2p)+p, & k=0,\\
                (\pi k)^{-1}(1-2p) \sin\left(2\pi k r\right), & k\in \N.
\end{array}
                                                                \right.
\end{equation*}
The order of the bifurcations is determined by the magnitude of the eigenvalues.
The first phase transitions correspond to the smallest positive eigenvalue  of $\bW$
if $\beta >0$, and the largest negative eigenvalue if $\beta<0$.

 Fig.~\ref{f.eigenval} shows several  $\mu_k$'s, with the
  four largest $(\mu_0 > \mu_1 > \mu_2 > \mu_3)$ and four smallest
  $(\mu_8 < \mu_9 < \mu_7 < \mu_{10})$
  eigenvalues are  highlighted in blue
and red respectively.

  These eigenvalues determine the bifurcating branches
  shown in Fig.~\ref{f.bif-sw}.
  The ordering of these eigenvalues explains the sequence, in which different harmonics emerge in the
  AFM regime:
  the first bifurcating solution is $\cos(16\pi x)$, followed by $\cos(18\pi x)$, $\cos(14\pi x)$, and
  $\cos(20\pi x)$.
\begin{figure}[h]
    \centering
    \includegraphics[width=\linewidth]{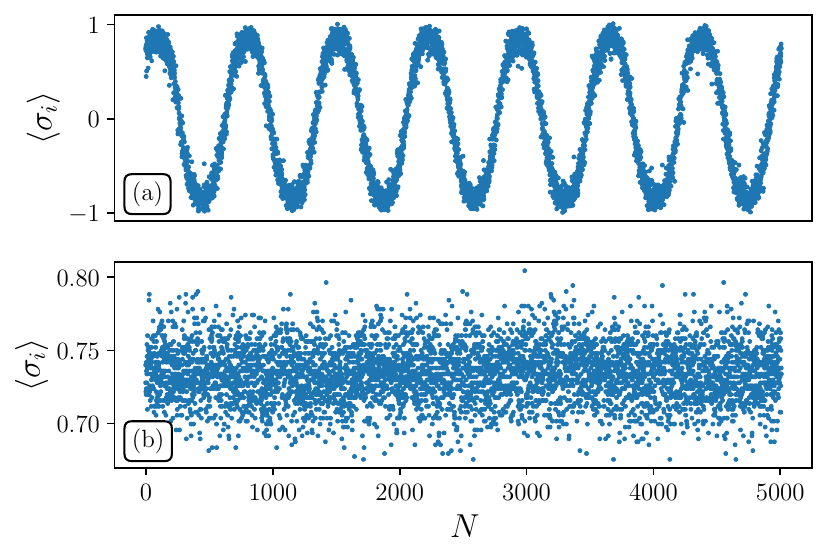}
    \caption{Spin configurations for: (a) AFM case ($J=-1$) slightly below $T=J\mu_8$, (b) FM case ($J=+1$) slightly below $T=J\mu_0$}
    \label{fig:stable-states}
  \end{figure}
  \begin{figure}[h]
    \centering
    \includegraphics[width=\linewidth]{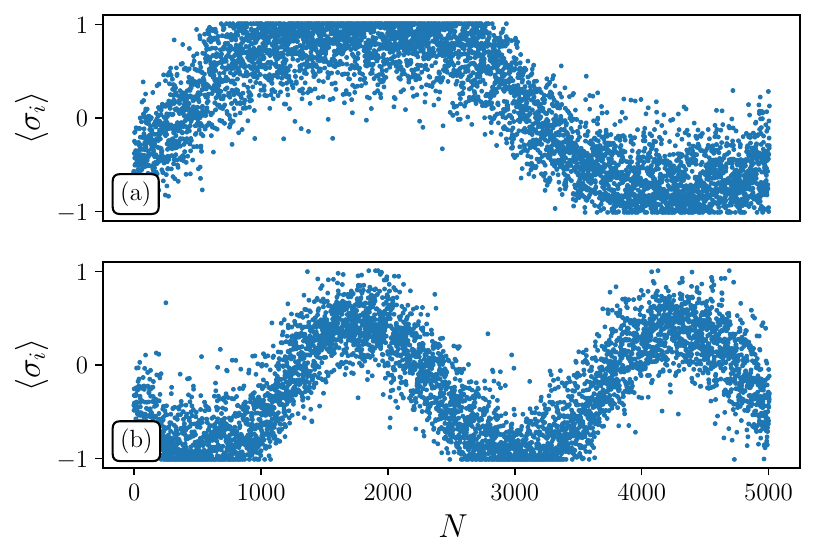}
    \caption{Spin configurations for: (a) FM case slightly below $T=J\mu_1$, (b) FM case slightly below $T=J\mu_2$}
    \label{fig:transient-states}
\end{figure}

\section{Monte-Carlo simulations}\label{sec.Monte-Carlo}

We performed Monte Carlo simulations using the Metropolis–Hastings algorithm~\cite[Ch.~4]{landau2021guide}. Metropolis algorithm constructs the Markov chain of states, where each new state $\bm{S}^{(k+1)}$ is generated from the current state $\bm{S}^{(k)}$ by a simple rule. This rule is based on the so-called \emph{proposal distribution} $g(\bm{S}'|\bm{S}^{(k)})$, where $\bm{S}'$ is the proposal for a new state and $\bm{S}^{(k)}$ is our current state. We accept this proposal state $\bm{S}'$ with acceptance probability,
\begin{equation}\label{eq:acceptance-prob}
    A(\bm{S}^{(k)}\to\bm{S}')=\min\left(1,\,\frac{p(\bm{S}')}{p(\bm{S})}\frac{g(\bm{S}'|\bm{S}^{(k)})}{g(\bm{S}^{(k)}|\bm{S}')}\right)
\end{equation}
Then, we generate a random number $a$ from the uniform distribution on $[0,1]$. If $a\leq A(\bm{S}^{(k)}\to\bm{S}')$, we accept our proposal state, i.e. $\bm{S}^{(k+1)}=\bm{S}'$. If $a>A(\bm{S}^{(k)}\to\bm{S}')$, we reject this change by keeping $\bm{S}^{(k+1)}=\bm{S}^{(k)}$. It is known that after a large enough number of steps we will obtain the desired Boltzmann-Gibbs distribution that delivers the global minima for free energy. The update rule is simple: we pick uniformly $i$-th from $N$ spins, so $g(\bm{S}'|\bm{S}^{(k)})=g(\bm{S}^{(k)}|\bm{S}')=N^{-1}$ and then we flip this spin. The acceptance probability is simply $\min(1,e^{-\beta\Delta E})$, where $\Delta E$ is the energy change after spin flip.

We have simulated the small-world graph with parameters $r = 0.1$, $p = 0.05$, and $N = 5000$ spins. Starting from a random (uniform) spin configuration, the system relaxes to the ground state corresponding to the principal bifurcating branch
at $\beta = \mu_0$ in the FM regime and $\beta = \mu_8$ in the AFM regime (see Fig.~\ref{f.bif-sw}). To identify metastable states associated with $\mu_k$, $k \notin {0,8}$, the system was initialized with $\operatorname{sign}\left(\bm{v}_k(x)\right)$, where $\bm{v}_k$ is the corresponding harmonic mode (see insets in Fig.~\ref{f.bif-sw}). For temperatures just above $T = J\mu_k$, the system remains for an extended period near the corresponding eigenstate before drifting away and eventually relaxing to the ground state (Fig.~\ref{fig:transient-states}). These simulations suggest that the coexistence of multiple solutions at low temperatures gives rise to metastable behavior. As $T \to 0$, the number of coexisting solutions of Eq.~\eqref{clim} can become arbitrarily large.

\begin{figure}
    \centering
    \includegraphics[width=\linewidth]{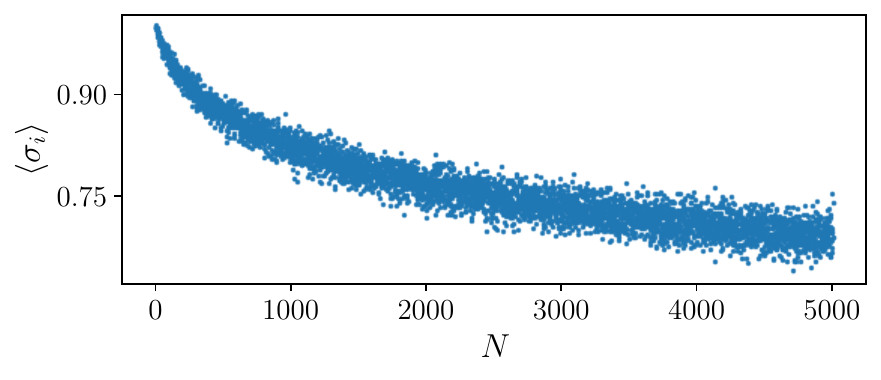}
    \caption{Spin configuration for power-law graph with $\alpha=0.2$ slightly below $T=J/(1-2\alpha)$}
    \label{fig:pl-FM-stable}
\end{figure}

For the power law graph, we set $\alpha=0.2$ and simulate with $N=5000$ spins below the critical temperature $T_c=J/(1-2\alpha)$. The spin configuration is shown at Fig.~\ref{fig:pl-FM-stable}.

\section{Discussion}\label{sec.discuss}

The Ising model of interacting spins has played a central role in statistical physics and its many
applications \cite{FriVel-StatMech}.
Within this framework, exactly solvable mean-field models, such as
the Curie-Weiss and Sherrington-Kirkpatrick models, have provided important insights into the equilibrium
statistical mechanics of interacting spin systems. In the mean-field setting, the Hamiltonian can
be expressed in terms of suitable macroscopic variables, such as the magnetization,
which, together with the large deviations formalism (cf.~\cite{Ellis-LDs}), leads to substantial simplifications and ultimately
to exact characterizations of thermodynamic behavior and phase transitions.

At the same time, the all-to-all coupling underlying classical mean-field models is 
physically unrealistic. Importantly, it obscures effects arising from the underlying
structural organization and heterogeneity of spin-spin interactions.

In this paper, we have shown that the use of graphons in the Ising model allows one to retain the
benefits of the mean-field formalism while, at the same time, incorporating nontrivial network
connectivity. This, in turn, leads to rich thermodynamic behavior, including multiple
phase transitions, coexistence of ferromagnetic and antiferromagnetic phases, and metastability.

For simplicity, in the present study we restricted ourselves to couplings of the same sign.
Clearly, one can also incorporate interactions with both positive and negative couplings,
as in the Sherrington-Kirkpatrick model, which may make graphon-based models
relevant for the study of spin glasses. Another aspect that we have only briefly touched
upon is metastability. The coexistence of steady states in \eqref{clim} may lead to
metastable behavior, as already confirmed by our Monte-Carlo simulations.

Another source of multistability was pointed out in \cite{MedPel2024}. It was observed that,
while the solutions of the continuum limit are translation invariant, this invariance
breaks down upon discretization, yielding multiple steady-state solutions for the discrete
system on random graphs (see Fig.~5 in \cite{MedPel2024}). Consequently, each branch bifurcating from the ground state
of the continuum system gives rise to multiple solutions of the discrete finite-size system,
thereby providing a mechanism for multistability.
These questions will be addressed in future work.

Using graphons, one can naturally incorporate a wide variety of network topologies into the Ising
model. While the classical theory of graph limits is formulated for dense networks, it extends
naturally to sparse networks with unbounded degree \cite{borgs2019Lp1}. Moreover, by using
techniques of
nonlocal-to-local approximation (cf.~\cite{PauTre2025}), one can further study physically relevant
models with
local interactions. Likewise, the graphon based
approach can be exteded to cover self-similar networks \cite{Med2026}.

Graphons, originally motivated by problems in graph theory and combinatorics, have rapidly
found applications in interacting particle systems, leading to remarkable progress
in the analysis
of dynamical processes on networks that, until recently, could only be studied numerically.
The study of phase transitions in the Kuramoto model of coupled phase oscillators provides
a vivid illustration of the effectiveness of the graphon-based approach to the analysis
of complex networks \cite{Med2014a,KVMed2018,ChiMed19a,MedMiz22,ChiMed22,CMM23}.
The same reasons that made graphons so successful in the study of dynamical models on
networks also make them promising tools for the analysis of spin systems. The results of the
present work clearly demonstrate the potential of
graphons for obtaining rigorous analytical results for spin models.

\textbf{Acknowledgements}--
AA and GSM thank Jaime Cisternas for suggesting \verb|BifurcationKit|
and for sharing his code.
AA acknowledges discussions with Dmitriy Lyubshin and Mauro Mariani on Monte Carlo simulations.
The research of AA is supported by Basic Research Program at the HSE University and by
the Foundation for the Advancement of Theoretical Physics and Mathematics ``BASIS''
(grant No. 25-1-4-2-1). The work of GSM was supported in
part by the NSF (DMS-2406941).
Numerical simulations were performed at the HSE University. 

AA and GSM designed the study. GSM analyzed the phase transitions. AA performed the
numerical bifurcation analysis and Monte-Carlo simulations.
Both authors interpreted the results and reviewed the manuscript.

The authors declare they have no conflicts of interest.

\bibliographystyle{apsrev4-2}
%

\end{document}